\documentclass[review]{elsarticle}

\usepackage[legalpaper, margin=1in]{geometry}

\usepackage{graphicx}
\usepackage{amssymb}
\usepackage{lineno}
\usepackage{lineno,hyperref}
\usepackage{amsmath}
\modulolinenumbers[5]
\usepackage{subfig}
\journal{Ocean Engineering}

\begin{document}
\begin{frontmatter}


\title{Experimental and numerical investigation of the hydrodynamic characteristics of Autonomous Underwater Vehicles over sea-beds with complex topography}



\author{A. Mitra} \author{J.P. Panda\fnref{myfootnote}} \author{H.V. Warrior}
\address{Department of Ocean Engineering and Naval Architecture,
IIT Kharagpur, India}

\fntext[myfootnote]{corresponding author: jppanda@iitkgp.ac.in (Dr. J. P. Panda)}

\begin{abstract}
Improved designs for Autonomous Underwater Vehicles (AUV) are becoming increasingly important due to their utility in academic and industrial applications. However, a majority of such testing and design is carried out under conditions that may not reflect the operating environment of shallow water AUVs. This may lead to imprecise estimations of the AUV's performance and sub-optimal designs. This article presents experimental and numerical studies carried out in conjunction, to investigate the hydrodynamic characteristics of AUV hulls at different Reynolds numbers over sloped channel-beds. We carry out experiments to measure the velocity field and turbulent statistics around the AUV with quantified uncertainty. These are contrasted against corresponding flat bed experiments to gauge the effect of test bed topography on AUV performance. The experimental data was used to validate Reynolds Stress Model predictions. Hydrodynamic parameters such as drag, pressure and skin friction coefficients were predicted from the RSM simulations at different test bed slopes, angles of attack and drift angles of the AUV hull, to analyze the hydrodynamic performance of the AUV. The results presented in this article offer avenues for design improvement of AUVs operating in shallow environments, such as the continental slope and estuaries. 
\end{abstract}


\end{frontmatter}

%\linenumbers

\section{Introduction}
\label{S:1}
The exploration of ocean and to interpret its underwater behavior is of importance in today’s world. To examine the ocean bed, convenient accessories are required to minimize the presence of human operators underneath of ocean. Autonomous Underwater Vehicles (AUV) are automated vehicles that are capable of underwater locomotion. These have applications in a variety of fields. For example, in the commercial field, oil and sub-sea drilling companies use AUVs for the purpose of checking the appropriate oceanic area where drilling may be optimally beneficial. For research and exploration, AUVs are used to track reefs and other life-forms that exist underwater. Additionally, AUVs have found applications in military and academic fields as well, besides others\cite{SAHOO2019145}. 

The initial AUV designs were just modifications on existing designs of submersible torpedoes\cite{blidberg2001development}. However, with the new found applications of these vehicles, focused investigation into their optimal design have become more critical. For instance, there is a demand for AUVs that can execute missions of the order of weeks and months. This requires that the AUV design be as optimal as possible, necessitating a careful analysis of the hydrodynamic performance of the AUV structure. Several experimental and numerical  studies are available in the literature to analyze the hydrodynamic parameters around AUVs \cite{mansoorzadeh2014investigation,jagadeesh2009experimental,saeidinezhad2015experimental,javadi2015experimental,hayati2013study,papadopoulos2015hydrodynamic,da2017numerical,jian2016numerical,rattanasiri2015numerical,leong2015numerical,gafurov2015autonomous,tyagi2006calculation,AMIRI2019192}.
Mansoorzadeh and Javanmard \cite{mansoorzadeh2014investigation} have studied the effect of free surface on drag and lift coefficients of AUV at different submergence depths and compared the experimental data with the computational fluid dynamics (CFD) simulations. It was observed that hydrodynamic coefficients were very much responsive to the submergence depth and AUV speed. Jagadees et al.\cite{jagadeesh2009experimental} made experimental analysis on hydrodynamic force coefficients (axial, normal, drag, lift and pitching moment) at different angles of attack and Reynolds numbers. Saeidinezhad et al.\cite{saeidinezhad2015experimental} performed experimental analysis of the effect of Reynolds numbers on the pitch and drag coefficient of a submersible vehicle model. Javadi et al.\cite{javadi2015experimental} conducted experimental analysis of the effect of bow profiles on the resistance of the AUV in a towing tank at different Froude numbers and have studied the variability of the friction drag with Froude numbers . Salari and Rava \cite{salari2017numerical} numerically studied the hydrodynamics over AUV near the free surface with different turbulence models. The wave effects on AUV were analyzed at different depths from sea surface. Jagadeesh and Murali\cite{jagadeesh2005application} have studied the hydrodynamic forces on AUV hull forms using different two equation turbulence models. De Sousa et al.\cite{de2014numerical} have analyzed the turbulent flow and drag coefficient to optimize the AUV hull design. A relationship between geometric parameters and the drag coefficients for different hull geometry was established. Alvarez et al.\cite{alvarez2009hull} examined the wave resistance on the AUV operating near the surface. Wu et al.\cite{wu2014hydrodynamic} explored the hydrodynamics of an AUV approaching the dock at different speeds in a cone-shaped dock under the influence of ocean currents. Optimum shape of the dock to minimize the drag have been considered. Leong et al.\cite{leong2015quasi} analyzed the hydrodynamic interaction effects of AUV nearer to a moving submarine. Tyagi and Sen \cite{tyagi2006calculation} performed CFD simulations to calculate the transverse hydrodynamic coefficients of an AUV hull, which are important in maneuverability study of marine vehicles. More recently \cite{mitra2019effects} have studied the effects of free stream turbulence on the hydrodynamic characteristics of an AUV hull form.

With the improvement of computational facilities, AUV design is increasingly becoming a simulation based methodology, with Computational Fluid Dynamics (CFD) simulations replacing the reliance on traditional large scale experiments over complex test beds. The use of such computer driven simulations enables rapid and inexpensive testing of complex design iterations that might be expensive and time consuming to replicate and test in real life. This also enables the aggregation of data at higher accuracy and of statistics that may not be possible in conventional experiments.

In the context of flow over AUV designs, most such flows are turbulent and the designers use turbulence models to account for the effects of turbulence. Most researchers investigating AUV designs(\cite{jagadeesh2005application,tyagi2006calculation,sakthivel2011application,mansoorzadeh2014investigation}) have used two equation turbulence models for the flow analysis across the AUV. The two equation based turbulence models like $k-\epsilon$ and $k-\omega$ models offer moderate fidelity in predictions at low computational cost. However due to the assumptions made in the development of such two-equation models they are limited in the turbulence physics that they can replicate. These assumptions include the use of a constitutive relations (the eddy viscosity hypothesis) instead of a transport equation for the Reynolds stress anisotropy, the use of the gradient diffusion hypothesis for the transport terms, etc. These limitations make the two-equation models ill suited for AUV design. As an illustrative example, we can analyze the appearance of flow separation over the exterior of the AUV. Such flow separation results in increased drag over the AUV, particularly the pressure drag caused by the pressure differential between the front and rear exterior surfaces. A critical design consideration in the design of AUV is to minimize such flow separation. However, two-equation models are ill suited to predict the onset and extent of such flow separation. This occurs due to the eddy viscosity hypothesis that assumes the eigenvectors of the mean velocity gradient to be aligned with the eigenvectors of the Reynolds stresses. This assumption is substantially violated in regions of flow separation, leading to unsatisfactory performance of two-equation models in such regions. Numerous studies have shown that two-equation models are ill suited in predicting the onset and extent of flow separation. Accordingly, the recent emphasis in the turbulence modeling community is shifting towards Reynolds stress models(\cite{mishra2019linear,panda2017, panda2018representation,panda2018experimental,mishrathesis,mishra1,mishra2,mishra3,mishra4,mishra5,mishra6}. However, the use of such Reynolds stress models in AUV design is very limited as of yet. In this investigation, we focus on this alternate modeling approach. The experimental measurements are used to calculate turbulence parameters such as turbulence kinetic energy and Reynolds stresses. These are then used to validate the Reynolds Stress Model(RSM) prediction of flow field along the AUV hull. Thereafter, hydrodynamic parameters such as drag, pressure and skin friction coefficients are predicted from the RSM simulations at different wedge heights using this validated model. The angle of attack and drift angle of the AUV hull were also varied to analyze the hydrodynamic performance of the AUV.

\begin{figure}
\begin{centering}
\includegraphics[width=0.8\textwidth]{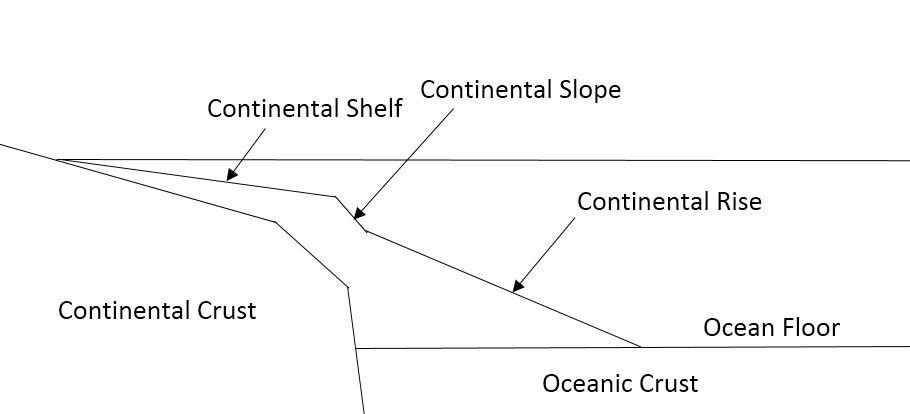}
\caption{Schematic illustration of the under water topology in a coastal zone \label{fig:1}}
\end{centering}
\end{figure}

A key shortcoming in the design of AUVs is the focus on operating environment. This is assumed to be the relatively flat ocean floor, evinced in the flat test beds used in AUV design. However, this operating environment is not a given. For instance, AUVs have to function exclusively in shallow waters\cite{shallow1,DU201476} (such as the mouths of rivers) or in coastal conditions\cite{shallow2} (such as near the continental shelf and slope). Testing the designs of such AUVs using conventional test beds leads to sub-optimal design for the operating conditions\cite{design1,design2}. AUVs operating under these conditions, such as at the mouth of the rivers, over Continental shelves (schematically outlined in figure \ref{fig:1}), etc, interact with complex strain fields because of complex slope of the sea-bed topography. There are little experimental or numerical results available in literature in which the detailed flow field near an AUV is studied over such underwater terrain. To address this need, we focus on experimental and numerical results over a sloped channel-bed. This simulates the operation of an AUV design for coastal operating environments, specifically over the continental shelf and slope. In this article we have conducted series of experiments at different Reynolds numbers and used an Acoustic Doppler Velocimeter to measure the flow field along the AUV hull. These are contrasted against the results from a corresponding flat test bed to demarcate the qualitative and quantitative aspects of the differences in hydrodynamic parameters arising due to differences in test bed topography. The results of turbulence stress components were utilized to validate the performance of a Reynolds stress model. After preliminary validation, several numerical simulations were performed to study the evolution of the drag, skin friction and pressure coefficients along the AUV hull. For the numerical simulations the wedge height was varied to change the bed-slope and study its effect on flow evolution along the AUV. The evolution of hydrodynamic parameters along the AUV were also studied by varying the angle of attack and drift angle at different wedge heights and Reynolds numbers, thus accounting for variation in operating environment of the AUV design.  

\begin{figure}
\centering
\subfloat[Schematic of the recirculating water tank]{\includegraphics[height=7cm]{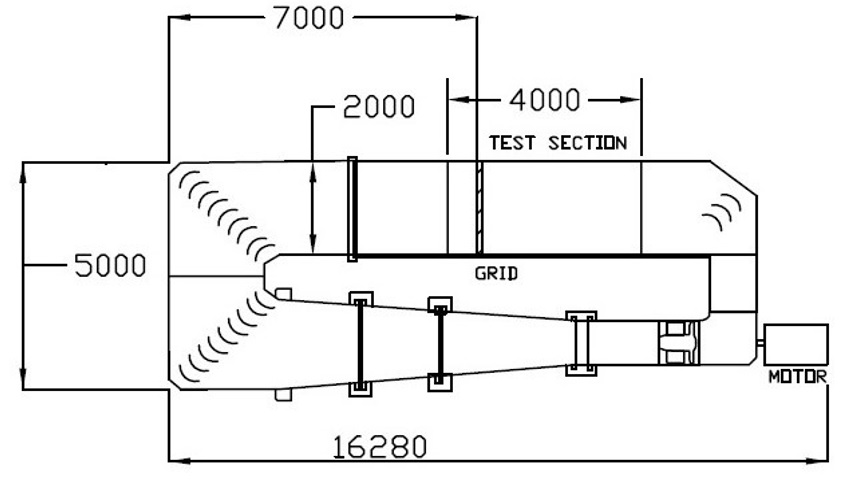}}\\
\subfloat[Experimental setup in the recirculating water tank, The detailed arrangement of AUV hull , wedge and the ADV. In this picture flow direction is from right to left.]{\includegraphics[height=7cm]{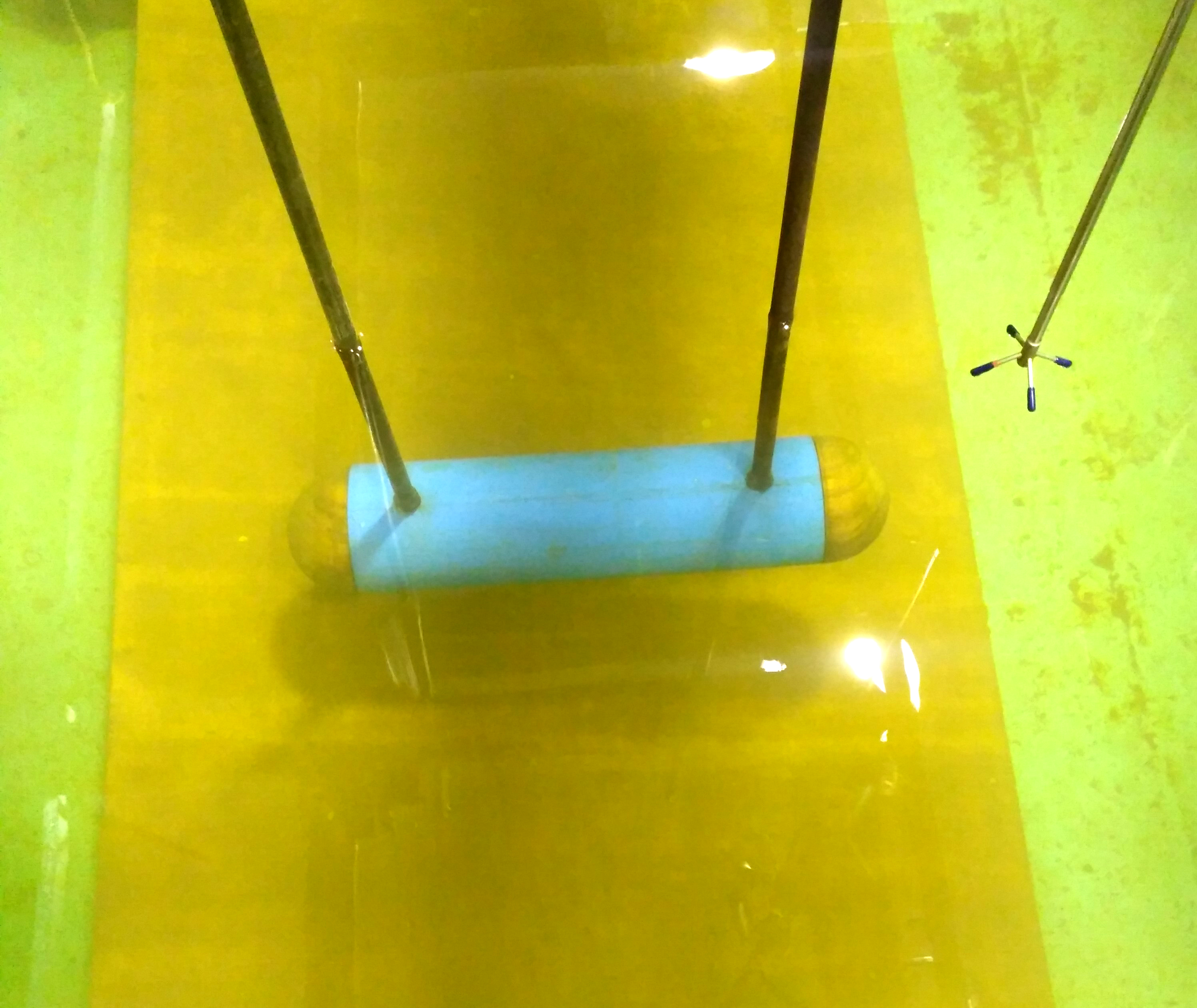}}\\
\subfloat[Detailed dimensions of AUV and wedge]{\includegraphics[height=7cm]{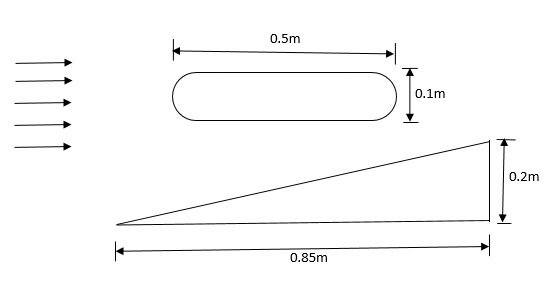}}
\caption{Experimental setup in the recirculating water tank, the arrow marks represents the flow direction, the AUV was fixed at the central line of water tank at a distance of 0.15 meter from the beginning of the wedge. Figure \ref{fig:2}a is reproduced from \cite{mitra2019effects} \label{fig:2}}
\end{figure}

\begin{figure}
\begin{centering}
\includegraphics[width=0.4\textwidth]{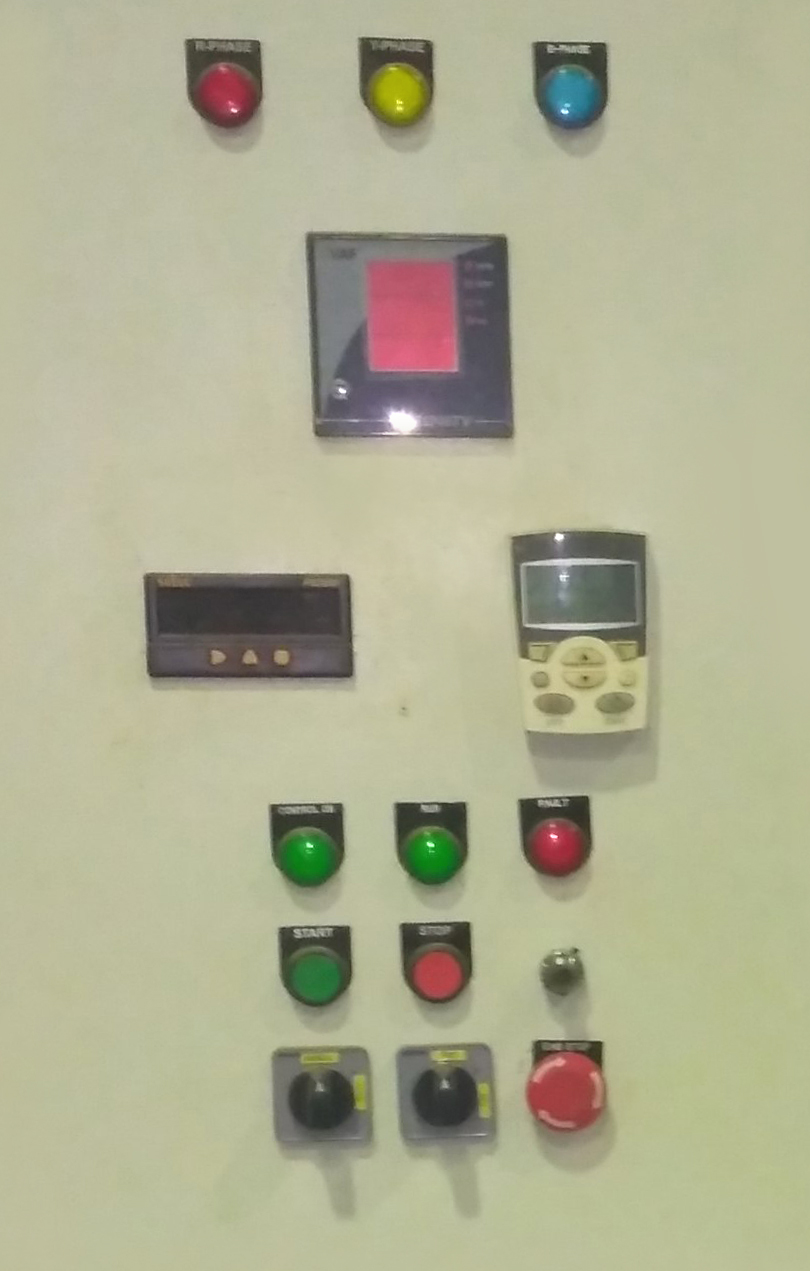}
\caption{The electrical control unit for controlling the RPM of the pump. \label{fig:3}}
\end{centering}
\end{figure}

\begin{figure}
\begin{centering}
\includegraphics[width=0.8\textwidth]{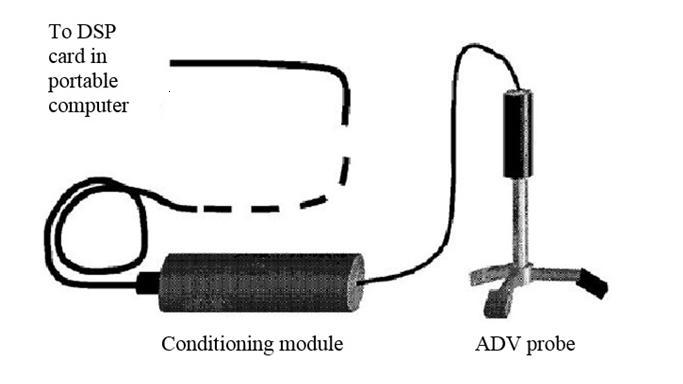}
\caption{Schematic diagram showing ADV probe and signal conditioning module, reproduced from \cite{panda2018experimental}. \label{fig:4}}
\end{centering}
\end{figure}
%%%%%%%%%%%%%%%%%%%%%%%%%%%%%%%%%%%%%%%%%%%%%%%%%%%%%%%%%%%%%%%%%%%%%%%%%%%%%%%%%%%%%%%%%%%%%%%%%%%%%%

\section{Experimental setup}
\label{S:2}
The experiments were conducted in a recirculating water channel at the department of Ocean Engineering and Naval Architecture,IIT, Kharagpur, India. The side walls and lower walls of the tank are made-up of transparent glass for proper flow visualization. The schematic of the recirculating water tank along with the detailed arrangement of sloped test bed(wedge) and AUV in the tank is shown in figure \ref{fig:2}. The detailed arrangement of the AUV and the wedge in the tank is shown in figure \ref{fig:2}b and the detailed dimensions are shown in figure \ref{fig:2}c. The the length and height of the wedge is 0.85 and 0.2 meter respectively. 

The water is recirculated by a pump, the pump rpm can be controlled through an electrical control unit as shown in figure \ref{fig:3}. By controlling the rpm of the pump the flow speed can be varied. For a water depth of $0.8$ meter a mean flow velocity up to $1 m/s$ is achievable. The water tank width and depth 2 meter and 1.5 meter respectively. All the experiments were conducted for a water depth of 0.8 meter in the water tank.

The AUV hull was fixed in the test section of the water tank just over the wedge. The arrow marks in figure\ref{fig:2}c represent the flow direction. The AUV hull has a cylindrical body with hemispherical ends as shown in figure \ref{fig:2}c. The length and diameter of the AUV hull is 0.5 and 0.1 meter respectively. The diameter of the hemispheres is equal to the diameter of the cylinder. The shape of the AUV hull in this work is based on the geometrical configuration investigated in \cite{mansoorzadeh2014investigation,mitra2019effects}. In this work we mainly have studied the effect of bed slope on the hydrodynamic performance of the AUV, so other accessories such as fins and propeller are not conneteced with the AUV hull.

The $x$-axis is the main flow direction, $y$ axis is the transverse direction and $z$ is the vertical direction. The measured velocities in transverse and vertical directions are comparatively smaller than the main flow direction velocities. $U, V$ and $W$ are the horizontal, transverse and vertical mean velocities in $x$, $y$ and $z$ directions respectively.

An ADV was used in our experiments, to collect the instantaneous velocities at different locations across the AUV, which is mostly suitable for flow measurements in  laboratory flumes and hydraulic models with higher sampling rate up to 200 Hz. At each location the ADV was fixed for five minutes, which is sufficient to obtain converged and stable instantaneous velocity data as reported in literature \cite{voulgaris}. A spatial and temporal resolution of $1 {cm^3}$ and $200 Hz$ respectively can be achievable through the ADV used in the experiment.

An ADV can measure instantaneous flow velocities at high sampling rates with very small sampling volume and works on the principle of Doppler shift. The three main components of ADV are a signal conditioning electronic module, sound receivers and sound emitter. The schematic of the acoustic doppler velocimeter is shown in figure \ref{fig:4}. More information on the ADV operation and working principle is available in \cite{panda2018experimental}. The ADV measurement errors are with in 1 percent as reported in \cite{voulgaris}.

The experiments were conducted for three different  volumetric  Reynolds numbers in the recirculating water tank, those varies form $Re_v=0.89\times 10^5$ to $Re_v=1.31\times 10^5$.  
\subsection{Data Analysis}
The data collected from the ADV were decomposed into mean and fluctuating velocities in x, y and z directions.

The stream-wise mean($U$) and fluctuating($u$) velocities were be calculated from the following formula:
\begin{equation}
U=1/n\sum_{i=1}^{n} U_i
\end{equation}

\begin{equation}
u=\sqrt{1/n\sum_{i=1}^{n} (U_i-U)^2}
\end{equation}
Similar formulas were employed to calculate the mean and fluctuating velocities in other two directions. 

The turbulent kinetic energy is defined as $k=\frac{1}{2}(\overline {u^2} + \overline {v^2} +\overline {w^2})$, which is the mean kinetic energy per unit mass in the fluctuating velocity field. Where v and w are the fluctuating velocities in transverse and vertical directions respectively.
%%%%%%%%%%%%%%%%%%%%%%%%%%%%%%%%%%%%%%%%%%%%%%%%%%%%%%%%%%%%%%%%%%%%
\section{Numerical modeling details}
\label{S:3}
The motion of the AUV under the free surface was modeled by solving the equation of conservation of mass and momentum for two phase flow:
\begin{equation}
\begin{split}
\partial_{t} {(\alpha_i\rho_i)}+\nabla.(\alpha_i\rho_i U)=0, \quad i=1,2,
\end{split}
\end{equation}

\begin{equation}
\begin{split}
\alpha_i=\frac{V_i}{V}, \quad i=1,2,
\end{split}
\end{equation}

\begin{equation}
\begin{split}
\sum_{i} \alpha_i=1,
\end{split}
\end{equation}

\begin{equation}
\begin{split}
\sum_{i} \nabla.(\alpha_i U)=0.
\end{split}
\end{equation}

\begin{equation}
\begin{split}
\partial_{t} {(\rho_{m} U)}+\nabla.(\rho_{m} U*U)=\nabla.(-P+\mu_m((\nabla U)+(\nabla U)^T)), \quad i=1,2,
\end{split}
\end{equation}
where $U$ is the velocity vector, $\alpha_i$ is the volume fraction of phase i, $V_i$ is the volume of phase i. $\rho_m$ and $\mu_m$ are the density and viscosity, respectively, and $P$ is the pressure acting on the flow. More information on the free surface modeling is available in \cite{ansys}.

We utilize Reynolds Stress Modeling based closures to account for the effects of turbulence in the CFD simulation. The building block of such models is the Reynolds stress transport equation, which has the form:
\begin{equation}
\begin{split}
\partial_{t} \overline{u_iu_j}+U_k \frac{\partial \overline{u_iu_j}}{\partial x_k}&=P_{ij}-\frac{\partial T_{ijk}}{\partial x_k}-\epsilon_{ij}+\phi_{ij},\\
\mbox{where},\\ 
P_{ij}=-\overline{u_ku_j}\frac{\partial U_i}{\partial x_k}-\overline{u_iu_k}\frac{\partial U_j}{\partial x_k}, & T_{kij}=\overline{u_iu_ju_k}-\nu \frac{\partial \overline{u_iu_j}}{\partial{x_k}}+\delta_{jk}\overline{ u_i \frac{p}{\rho}}+\delta_{ik}\overline{ u_j \frac{p}{\rho}}\\
,\epsilon_{ij}=-2\nu\overline{\frac{\partial u_i}{\partial x_k}\frac{\partial u_j}{\partial x_k}},& \phi_{ij}= \overline{\frac{p}{\rho}(\frac{\partial u_i}{\partial x_j}+\frac{\partial u_j}{\partial x_i})}\\
\end{split}
\end{equation}
$P_{ij}$ denotes the production of turbulence, $T_{ijk}$ is the diffusive transport, $\epsilon_{ij}$ is the dissipation rate tensor and $\phi_{ij}$ is the pressure strain correlation. The pressure fluctuations are governed by a Poisson equation:
\begin{equation}
\frac{1}{\rho}{\nabla}^2(p)=-2\frac{\partial{U}_j}{\partial{x}_i}\frac{\partial{u}_i}{\partial{x}_j}-\frac{\partial^2 u_iu_j}{\partial x_i \partial x_j}
\end{equation}

The fluctuating pressure term is split into a slow and rapid pressure term $p=p^S+p^R$. Slow and rapid pressure fluctuations satisfy the following equations
\begin{equation}
\frac{1}{\rho}{\nabla}^2(p^S)=-\frac{\partial^2}{\partial x_i \partial x_j}{(u_iu_j-\overline {u_iu_j})}
\end{equation}
\begin{equation}
\frac{1}{\rho}{\nabla}^2(p^R)=-2\frac{\partial{U}_j}{\partial{x}_i}\frac{\partial{u}_i}{\partial{x}_j}
\end{equation}
It can be seen that the slow pressure term accounts for the non-linear interactions in the fluctuating velocity field and the rapid pressure term accounts for the linear interactions. A general solution for $\phi_{ij}$ can be obtained by applying Green's theorem to equation (7):
\begin{equation}
\phi_{ij}=\frac{1}{4\pi}\int_{-\infty}^{\infty} \overline{\frac{\partial{u}^*_k}{\partial{x_l^*}}\frac{\partial{u}^*_l}{\partial{x_k^*}}(\frac{\partial{u}_i}{\partial{x_j}}+\frac{\partial{u}_j}{\partial{x_i}})}
+2G_{kl}\overline{\frac{\partial{u}^*_l}{\partial{x_k^*}}(\frac{\partial{u}_i}{\partial{x_j}}+\frac{\partial{u}_j}{\partial{x_i}})}\frac{dVol^*}{|{x_n-x_n^*}|}
\end{equation}
The volume element of the corresponding integration is $dVol^*$. Instead of an analytical approach, the pressure strain correlation is modeled using rational mechanics approach. The rapid term can be modeled by assuming the length scale of mean velocity gradient is much larger than the turbulent length scale and is written in terms of a fourth rank tensor \cite{pope2000}
\begin{equation}
\phi_{ij}^R=4k\frac{\partial{U}_l}{\partial{x_k}}(M_{kjil}+M_{ikjl})
\end{equation}
where, 
\begin{equation}
M_{ijpq}=\frac{-1}{8\pi k}\int \frac{1}{r} \frac {\partial^2 R_{ij}(r)}{\partial r_p \partial r_p}dr
\end{equation}
where, $R_{ij}(r)=\langle u_i(x)u_j(x+r) \rangle$

For homogeneous turbulence the complete pressure strain correlation can be written as
\begin{equation}
\phi_{ij}=\epsilon A_{ij}(b)+kM_{ijkl}(b)\frac{\partial\overline {v}_k}{\partial{x_l}}
\end{equation}
The most general form of slow pressure strain correlation is given by
\begin{equation}
\phi^{S}_{ij}=\beta_1 b_{ij} + \beta_2 (b_{ik}b_{kj}- \frac{1}{3}II_b \delta_{ij})
\end{equation}
Established slow pressure strain correlation models including the models of \cite{rotta1951} and \cite{ssmodel} use this general expression. Considering the rapid pressure strain correlation, the linear form of the model expression is
\begin{equation}
\frac{\phi^{R}_{ij}}{k}=C_2 S_{ij} +C_3 (b_{ik}S_{jk}+b_{jk}S_{ik}-\frac{2}{3}b_{mn}S_{mn}\delta_{ij})+  
C_4 (b_{ik}W_{jk} + b_{jk}W_{ik})
\end{equation}
Here $b_{ij}=\frac{\overline{u_iu_j}}{2k}-\frac{\delta_{ij}}{3}$ is the Reynolds stress anisotropy tensor, $S_{ij}$ is the mean rate of strain and $W_{ij}$ is the mean rate of rotation. Rapid pressure strain correlation models like the models of \cite{mishra6} use such a expression, linear in the Reynolds stress anisotropy. In the simulations outlined in this article, we utilize the closure model of \cite{mishra6} for the rapid pressure strain correlation. For the slow pressure strain correlation, we use the model of \cite{rotta1951}. This combination of slow and rapid pressure strain closures, both of which use closure expressions linear in the Reynolds stresses have been tested extensively in prior literature\cite{mishra6}.

The wall reflection term, redistributes the normal stress near the wall. The term is formulated such that it damps the component of Reynolds stress which is perpendicular to the wall and enhances the Reynolds stress parallel to the wall as reported in \cite{pope2000}. The wall reflection has both slow and rapid contributions and can be written as:
\begin{equation}
\begin{split}
\phi_{ij,Sw}=0.5\frac{\epsilon}{k}\bigg(\overline{u_{k}u_{m}}n_{k}n_{m}\delta{ij}-1.5\overline{u_{i}u_{k}}n_{j}n_{k}-1.5\overline{u_{j}u_{k}}n_{i}n_{k}\bigg)\frac{c_{l}k^{3/2}}{\epsilon d}\\
\phi_{ij,Rw}=0.3\phi_{km,R}\frac{\epsilon}{k}\bigg(n_{k}n_{m}\delta{ij}-1.5\phi_{ik,R}n_{j}n_{k}-1.5\phi_{jk,R}n_{i}n_{k}\bigg)\frac{c_{l}k^{3/2}}{\epsilon d}
\end{split}
\end{equation}
where, $n_k$ and $x_k$ are the components of the unit normal to the wall, $d$ is the normal distance to the wall\cite{gibson1978ground,ansys}.

The computational fluid dynamics(CFD) simulations were performed using the ANSYS Fluent solver \cite{ansys}. The full set of incompressible Navier-Stokes equations were solved utilizing the control volume approach. The coupling of pressure and velocity were made using the SIMPLE (Semi-Implicit Method for Pressure Linked Equations) scheme . The GAMBIT meshing software was used to generate spatial meshes with tetrahedral elements. The mesh independent study was performed with three meshes of increasing resolution. The cell size was varied as, $0.8 \times 10^6$, $1.5 \times 10^6$ and $2.2 \times 10^6$ respectively. Since we have used RSM model for our simulations, the $y^{+}$ value is taken as $70$. Considering that, at three different Reynolds numbers, the first layer thickness was calculated as $0.37m$, $0.3m$ and $0.26m$ respectively for the AUV hull. Inflation up to 5 layers having a growth rate of 1.2 was used in generating mesh for the domain. From the numerical simulations it was observed that the third mesh with 2.2 Million cells better predicts the turbulence kinetic energy and Reynolds stress components better matches with experimental results, so for all the simulations we have used the mesh with 2.2 million cells. The inlet and outlet of the water tank were modeled as a velocity inlet with a uniform inflow and a pressure outlet respectively.

%%%%%%%%%%%%%%%%%%%%%%%%%%%%%%%%%%%%%%%%%%%%%%%%%%%%%%%%%%%%%%%%%
\section{Results and discussions}
\label{S:4}
\subsection{Experimental results}
The measurement of instantaneous three dimensional velocities were taken along the length of the AUV at a distance of 0.05 m from the side walls and at six equidistant points starting from the beginning towards the end of the AUV hull. The mean velocity measurement is non-dimensionalised by U (the time averaged free stream mean velocity), the Reynolds stresses and turbulent kinetic energy are non-dimensionalised by $U^2$.

\begin{figure}
\centering
\subfloat[ ]{\includegraphics[height=8cm]{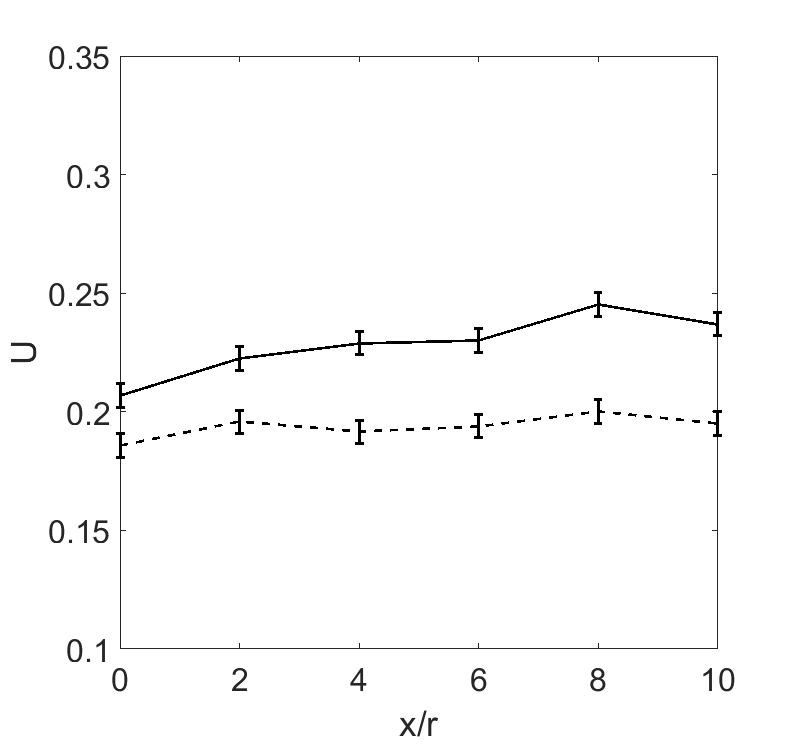}}
\subfloat[ ]{\includegraphics[height=8cm]{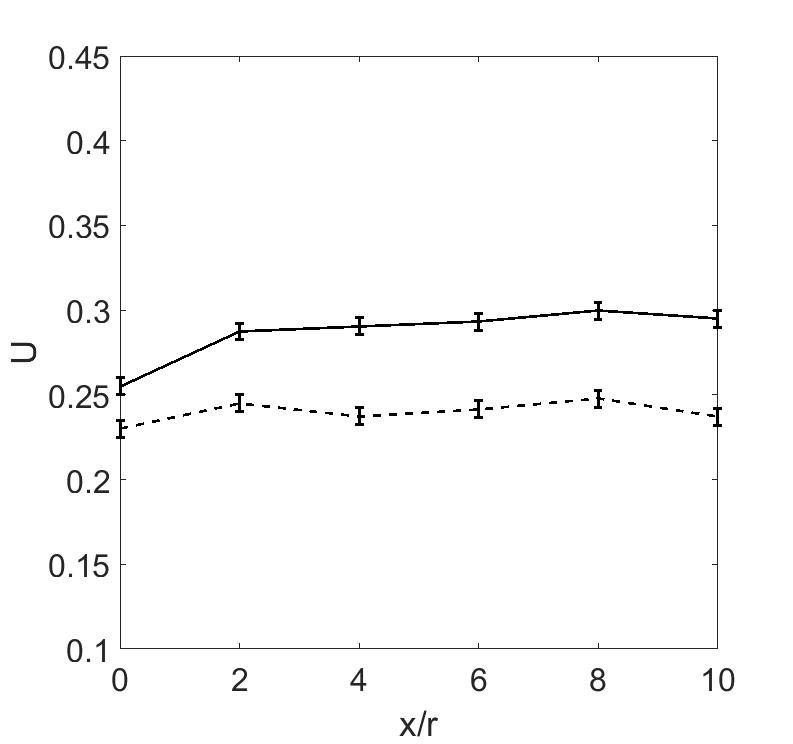}}\\
\subfloat[ ]{\includegraphics[height=8cm]{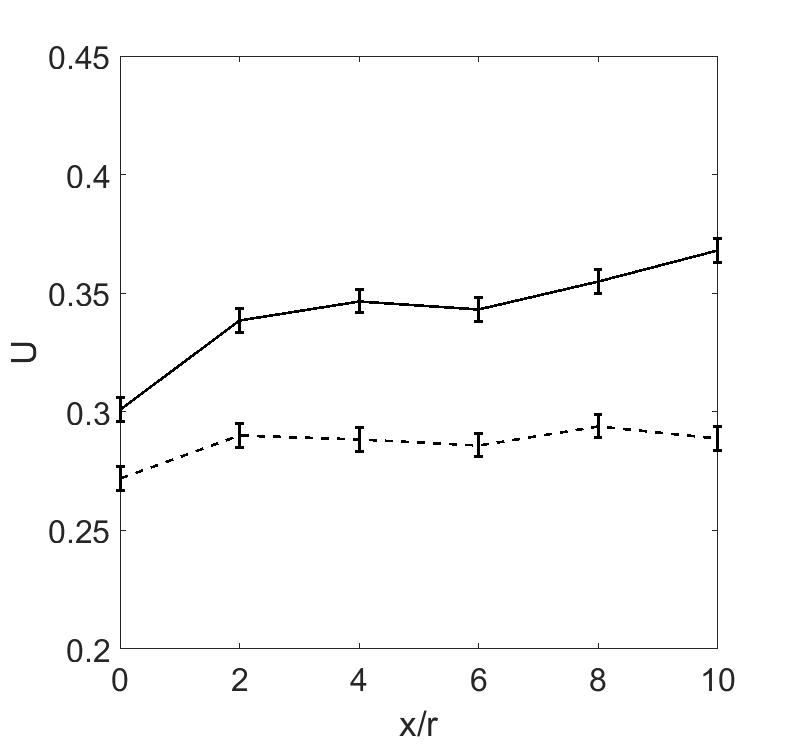}}
%\subfloat[ ]{\includegraphics[height=8cm]{fig/4b.jpg}}
\caption{Evolution of mean velocity along the AUV hull for a) $Re_v=0.89\times 10^5$ b) dashed line $Re_v=1.11\times 10^5$ c) $Re_v=1.31\times 10^5$. The dashed line and solid line represent the non-dimensional instantaneous velocity for the case of the flat test bed and sloped test bed cases respectively. \label{fig:5}
}
\end{figure}

The figure \ref{fig:5} shows the evolution of mean velocity along the AUV hull for three volumetric Reynolds numbers ($Re_v={\rho U \nabla^{\frac{1}{3}}}/{U}$). We contrast the mean velocity measurements over the sloped test bed with the corresponding measurements carried out over a conventional flat test bed as a datum for reference. The dashed line shows the velocity evolution for the flat test bed case and the solid line shows the evolution of velocity along the AUV fitted over the sloped test bed. A sharp increase in velocity is observed at all points across the AUV, in the case of the sloped test bed. For the $Re_v=1.31\times 10^5$ case, the drag force ($F_D$) over the AUV in operation over the sloped test bed was more than twice that over the AUV operating over the flat test bed. It should be emphasized that the slope of the test bed is a relatively low $13^o$, comparable to that found at the continental slope. Under such a increase in drag, the performance of an AUV designed using flat test beds would be significantly deteriorated. This would have a cascading effect on the range of operation of the AUV and its operation life.

\begin{figure}
\centering
\subfloat[Flat bed]{\includegraphics[height=8cm]{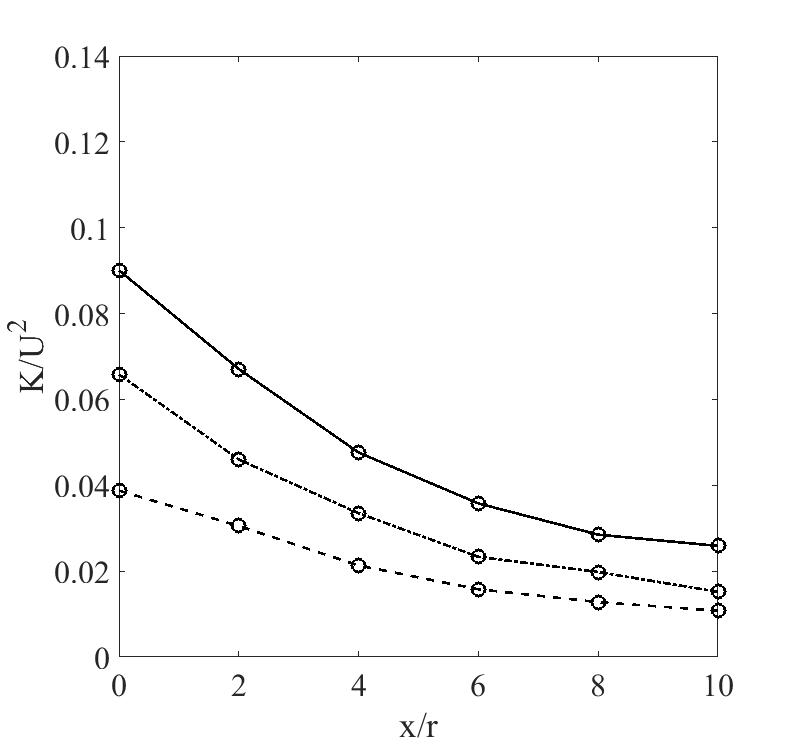}}
\subfloat[Sloped bed]{\includegraphics[height=8cm]{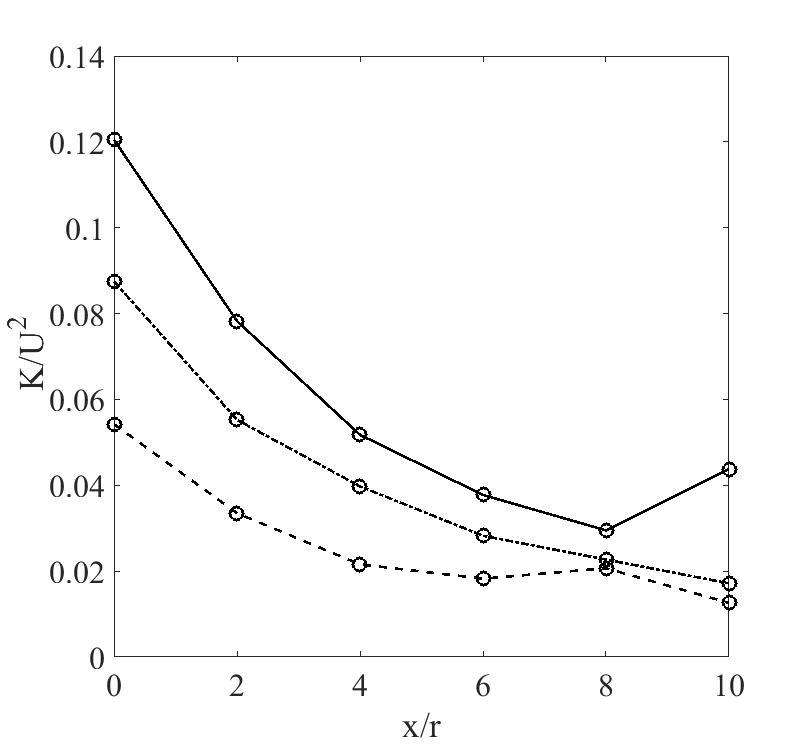}}
\caption{Evolution of turbulence kinetic energy along the AUV hull, Solid Line $Re_v=0.89\times 10^5$, dashed line $Re_v=1.11\times 10^5$ and dashed dot line for $Re_v=1.31\times 10^5$. \label{fig:6}}
\end{figure}

The variation of turbulence kinetic energy calculated from the velocity statistics are plotted in figure \ref{fig:6}b for three volumetric Reynolds numbers. Once again, we contrast the flow statistics over the sloped test bed case with that over the flat test bed case. On contrasting the figures, we observe that the qualitative trends in turbulent kinetic energy distribution over the surface of the AUV is similar for both the flat and the sloped test beds. However there is a significant increase in the value of the turbulent kinetic energy at the nose of the AUV for the sloped test bed case. This increased turbulent kinetic energy is reflective of higher turbulent intensity in the flow around the nose of the AUV. This higher turbulence intensity would lead to faster dissipation of momentum in this region, thus deteriorating AUV performance. 

\subsection{Validation of the numerical model}
\begin{figure}  
\centering
\subfloat[ ]{\includegraphics[height=8cm]{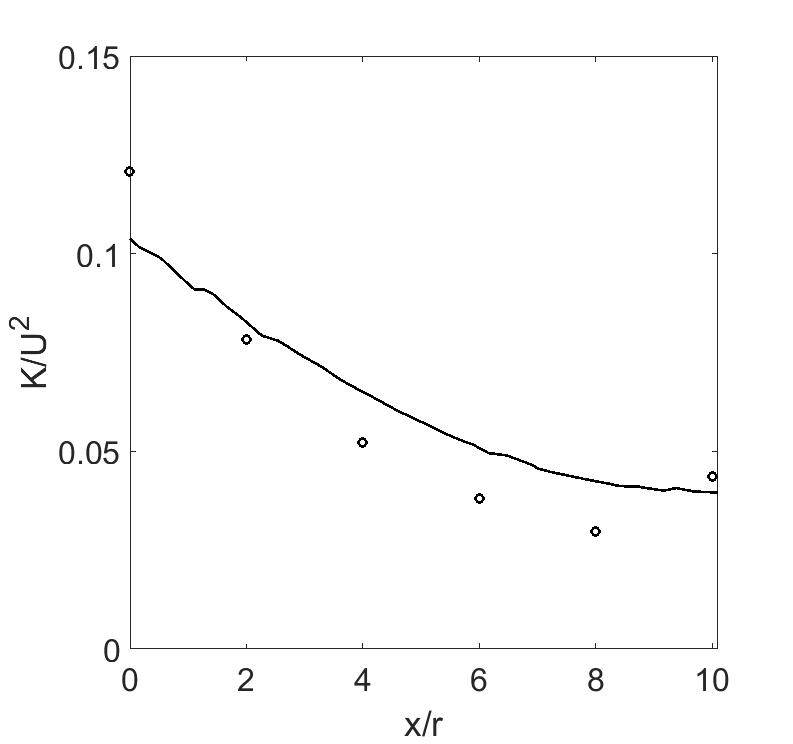}}\\
\subfloat[ ]{\includegraphics[height=8cm]{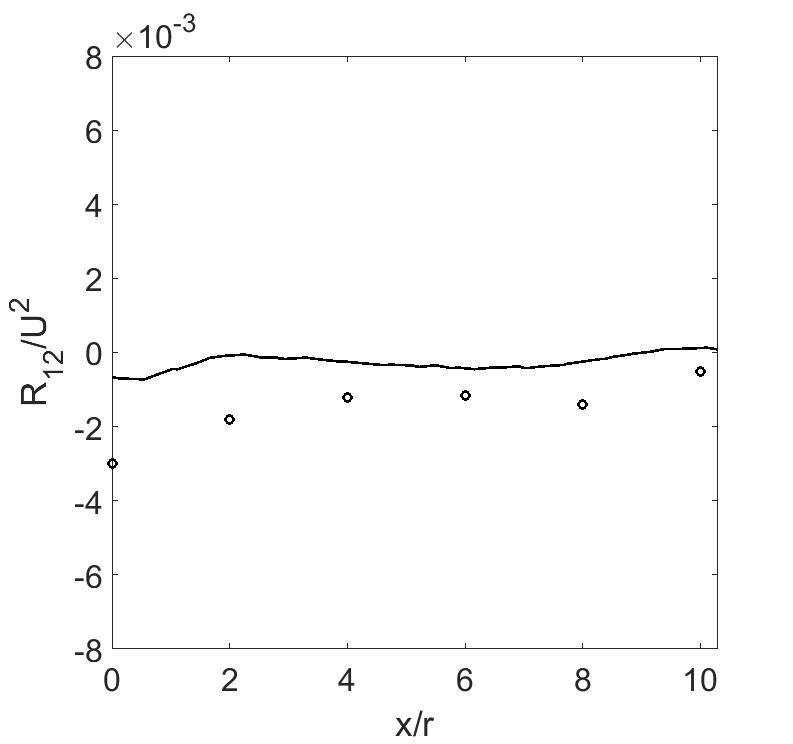}}
\subfloat[ ]{\includegraphics[height=8cm]{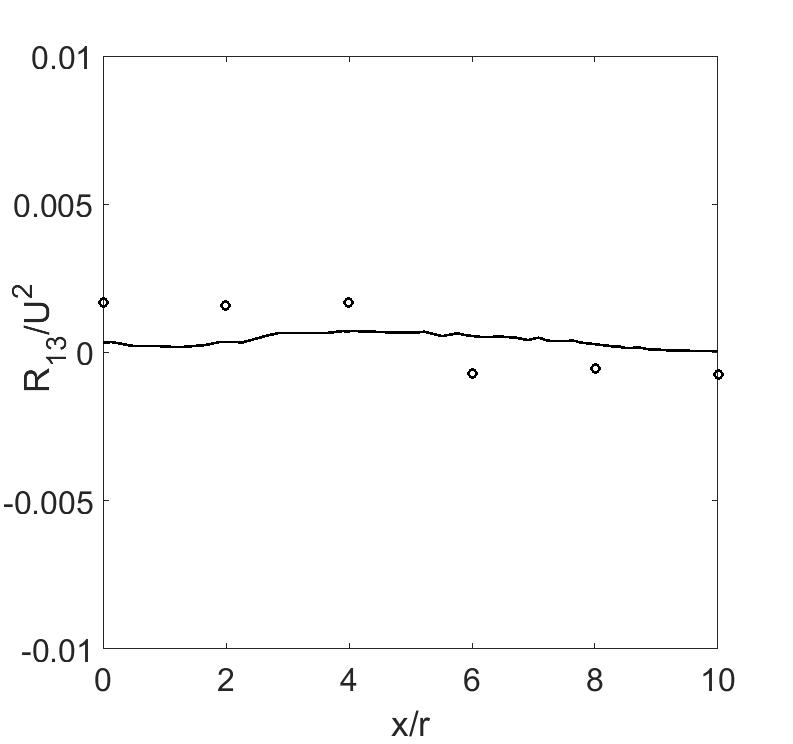}}
%\subfloat[ ]{\includegraphics[height=8cm]{fig/cp_validation.jpg}}
\caption{Comparison of Reynolds stress model predictions with the turbulence kinetic energy and  the components of Reynolds shear stress, 
a) evolution of turbulence kinetic energy b) evolution of shear component of Reynolds stress($R_{12}=\overline{uv}$), c) evolution of shear component of Reynolds stress($R_{13}=\overline{uw}$). The circled points represent the experimental results and the solid lines represent the Reynolds stress model predictions \label{fig:7}}
\end{figure}

For the validation of the numerical model, the experimental results for the case of flow past AUV in presence wedge is considered. The numerical simulations were conducted with the linear pressure strain model\cite{ansys}. Figure \ref{fig:7} shows the comparison of the Reynolds Stress model predictions of turbulence kinetic energy Reynolds shear stresses. From all the figures it is observed that the pressure strain correlation model with the wall reflection term well predicts the flow field. The wall reflection term accurately captures the modified pressure field in the proximity of the rigid AUV wall and impedes the transfer of energy from the stream wise direction to that normal to the wall as reported in \cite{mitra2019effects}.

%%%%%%%%%%%%%%%%%%%%%%%%%%%%%%%%%%%%%%%%%%%%%%%%%%%%%%%%%%%%%%%%%%%%
\subsection{Numerical results}
%%%%%%%%%%%%%%%%%%%%%%%%%%%%%%%%%%%%%%%%%%%%%%%%%%%%%%%%%%%%%%%%%%%%

The drag coefficient is given by:
 \begin{equation}
 C_d=\frac{F_d}{0.5\rho U^2A}
 \end{equation}
 Where, $F_d$ is the drag force acting on the cylinder, $A$ is the area of the external surface of the model.
 
The skin friction coefficient can be calculated as:
 \begin{equation}
 C_f=\frac{\tau_w}{0.5\rho{v^2}}
 \end{equation}
 where, $\tau_w$ is the skin shear stress on a surface, $v$ is the free stream velocity and the rest of the symbols have their usual meaning.
 The pressure coefficient is defined as:
 \begin{equation}
 C_p=\frac{p-p_\infty}{0.5\rho U^2}
 \end{equation}
 Where, p is the pressure at the point for which pressure coefficient is calculated.
 
\subsubsection{Effect of test bed slope variation on the hydrodynamic performance of the AUV}
\begin{figure}  
\centering
\subfloat[ ]{\includegraphics[height=8cm]{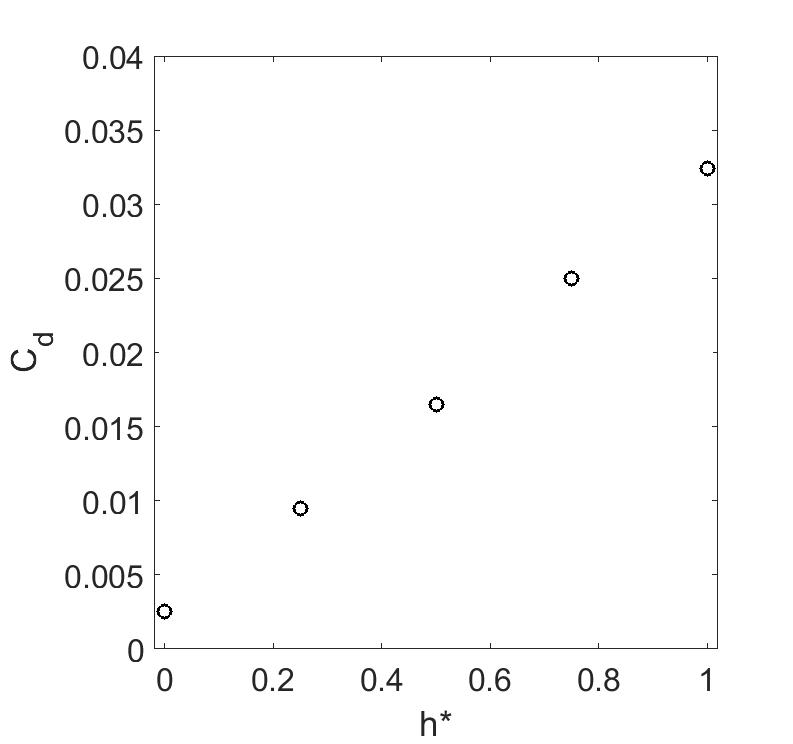}}
\subfloat[ ]{\includegraphics[height=8cm]{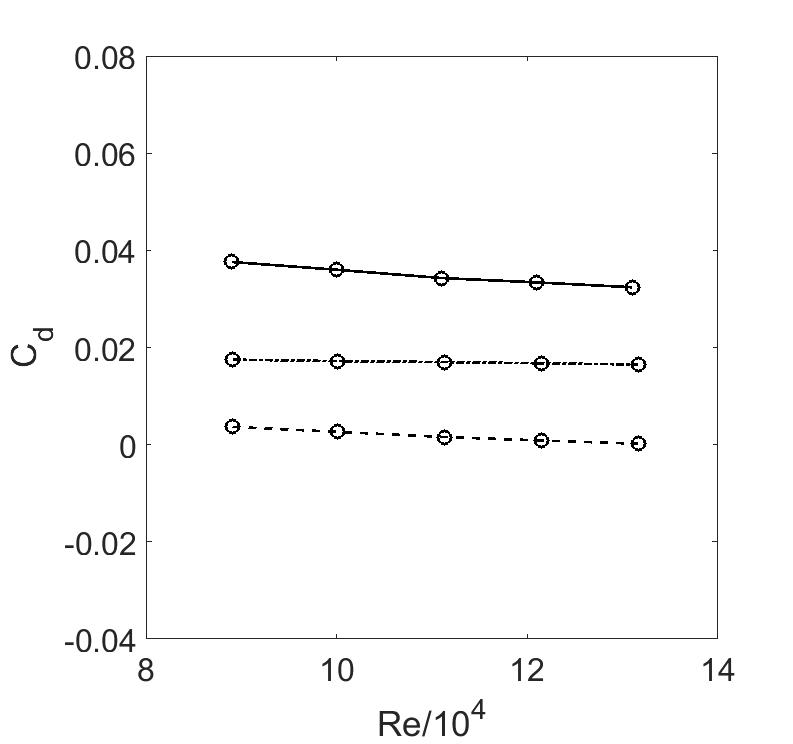}}
\caption{The variation of drag coefficient a) with wedge height for Reynolds number  $Re_v=1.31\times 10^5$ and b) with Reynolds number for different values of $h*$, dashed line, dashed-dot line and solid line represent $h^*=0$,1 and 2 respectively.
\label{fig:8}}\end{figure}

\begin{figure}  
\centering
\subfloat[ ]{\includegraphics[height=8cm]{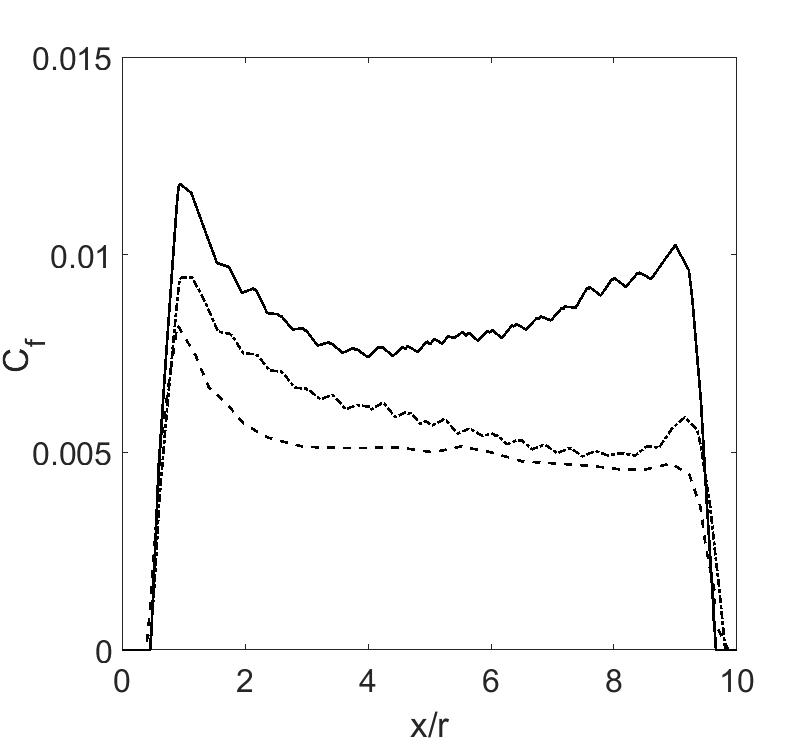}}
\subfloat[ ]{\includegraphics[height=8cm]{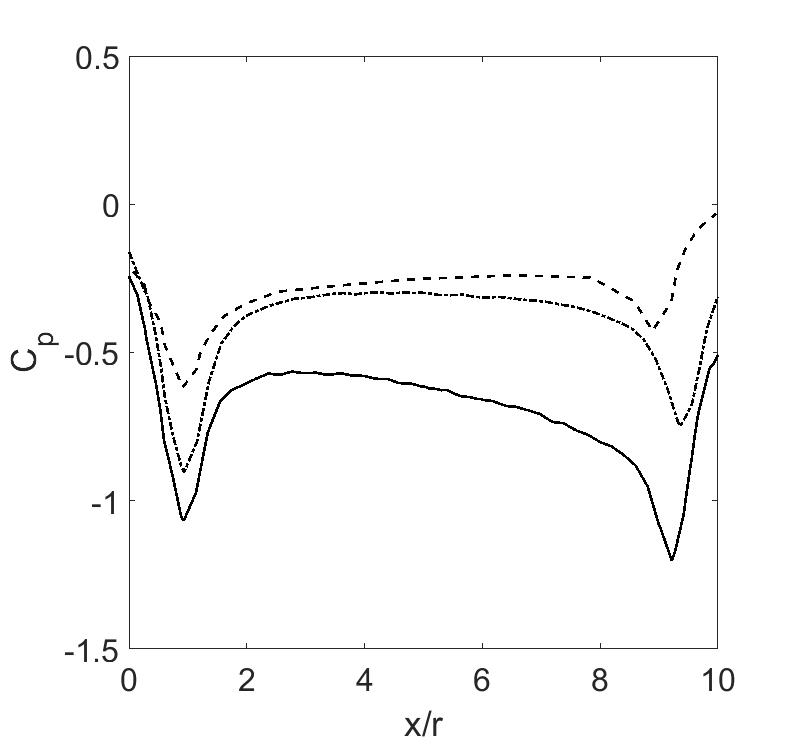}}
\caption{The effect of wedge height variation on skin friction and pressure coefficient of the AUV hull form. Dashed line,  dashed-dot line and solid line  represent $h^*=0$,1 and 2 respectively.
\label{fig:9}}
\end{figure}

%-----------------------------------------------------------
Figure \ref{fig:8} depicts the variation of drag coefficients of the AUV for different values of $h^*$. Here $h^*$ is the non dimensional wedge height. The wedge height is non-dimensionalized with the maximum wedge height. For our work the maximum wedge height is 0.2 meter, which is sufficient for contracting the flow and generating a pressure gradient field in the zone of operation of the AUV. Although we have conducted experiments only for the maximum wedge height, for the validation of the numerical model, for numerical simulations, we have varied the wedge height from a minimum of $h^*=0$ to $h^*=1$ with an increment of 0.25 to study the detailed evolution of hydrodynamic parameters along the AUV. Figure \ref{fig:8}a represent the evolution of the drag coefficient with the variation of the non dimensional wedge height for volumetric Reynolds number $Re_v=1.31\times 10^5$. A sharp increase in drag coefficient is observed with increase in $h^*$, this is because of the interaction of the pressure gradient field with the flow field of the AUV. In figure \ref{fig:8}b a detailed evolution of the drag coefficient is presented for different step height with variation in $Re_v$. For all three Reynolds numbers a decrease in drag coefficient is observed as for all the three values of wedge height. Figure \ref{fig:9} represent the skin friction and pressure coefficient evolution for three different values wedge height. It is observed that with increase in step height the skin friction coefficient increase over the AUV hull and a larger increase is observed for $h^*=1$. For pressure coeffcient a reverse trend observed in its evolution,i.e. the pressure coefficient is decreasing with increase in $h^*$.

%-----------------------------------------------------------
\subsubsection{Effect of angle of attack on the hydrodynamic performance of the AUV over the sloped channel-bed}
%-----------------------------------------------------------

\begin{figure}
\centering
\subfloat[ ]{\includegraphics[height=8cm]{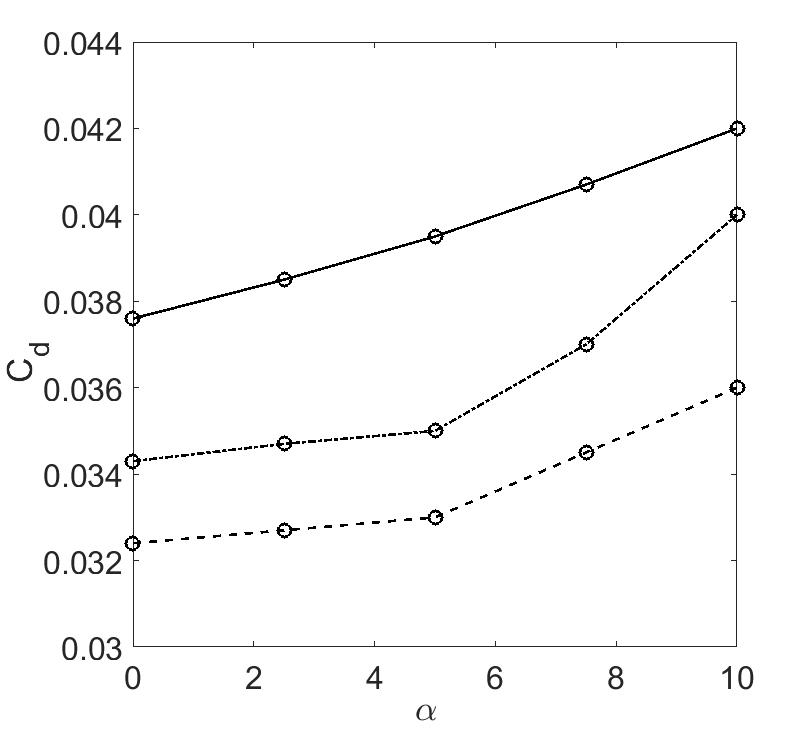}}
\subfloat[ ]{\includegraphics[height=8cm]{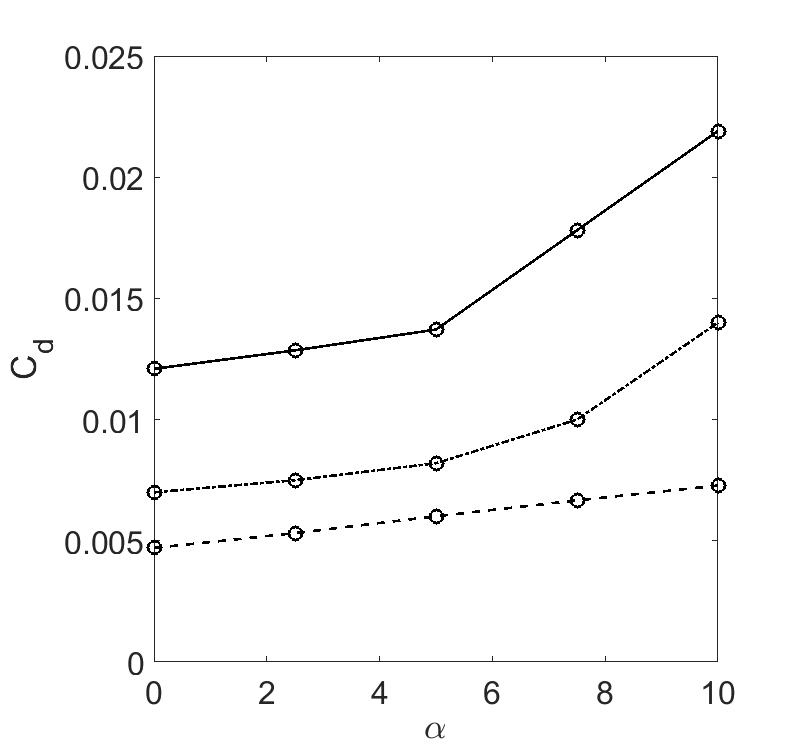}}
 \caption{The variation of drag coefficient for different values of AOA($\alpha$) a) with Reynolds number variation, dashed line, dashed-dot line and solid line represent $Re_v=1.31\times 10^5$, $Re_v=1.11\times 10^5$ and $Re_v=0.89\times 10^5$ respectively, b) with $h^*$ variation dashed line, dashed-dot line and solid line represent $h^*=0$, $h^*=0.5$ and $h^*=1$ respectively.
 \label{fig:10}}
 \end{figure}

\begin{figure}
\centering
\subfloat[ ]{\includegraphics[height=8cm]{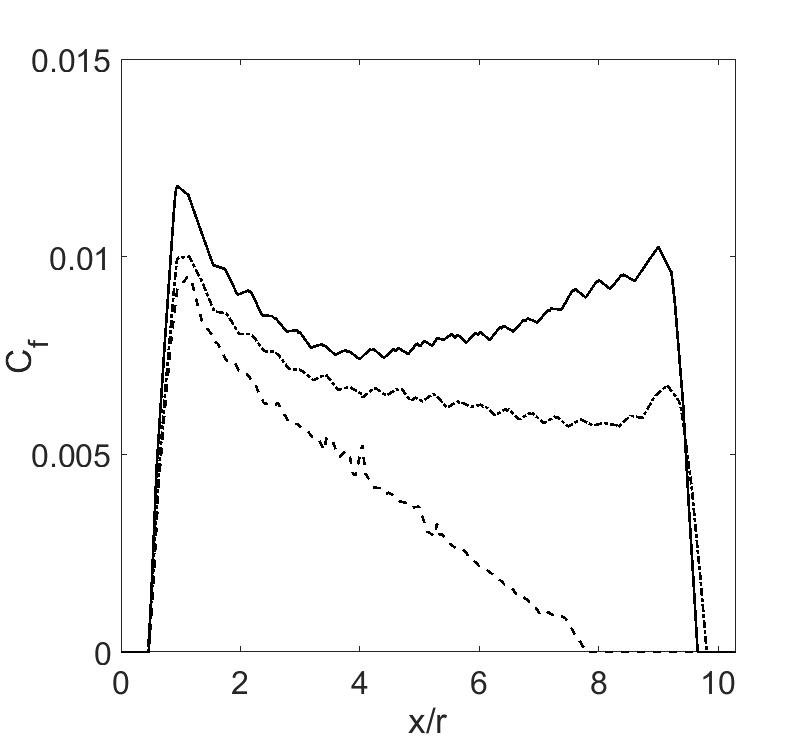}}
\subfloat[ ]{\includegraphics[height=8cm]{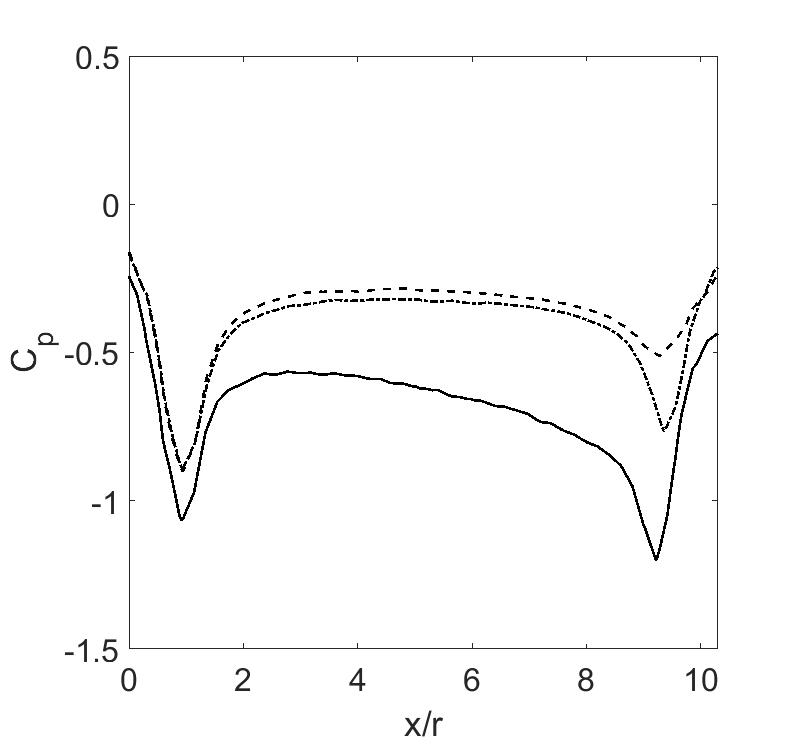}}
 \caption{The variation of skin friction and pressure coefficient for different values of AOA, dashed line, dashed-dot line and solid line represent AOA 0, 5 and 10 degrees respectively \label{fig:11}}
 \end{figure}

In Figure \ref{fig:10} the evolution of drag coefficient for different angles of attack is presented. Figure \ref{fig:10}a represent the the variation of drag coefficient with AOA for different volumetric Reynolds numbers. The dashed line is for the highest Reynolds number. It is observed that with increase in Reynolds number the drag coefficient decreases for all angles of attack. Figure \ref{fig:10}b represent the drag coefficient evolution with AOA for different values of wedge height. It is clear that with increase in wedge height drag coefficient increases for all values of AOA. The variation of skin friction and pressure coefficient with AOA is presented in \ref{fig:11}. For both the figures $h^*$ was taken as 1. An increase in skin friction coefficient is observed With increase in angle of attack. However a reverse trend is observed in the pressure coefficient evolution.

%-----------------------------------------------------------
\subsubsection{Effect of drift angle on the hydrodynamic performance of the AUV over the sloped channel-bed}
\begin{figure}
\centering
\subfloat[ ]{\includegraphics[height=8cm]{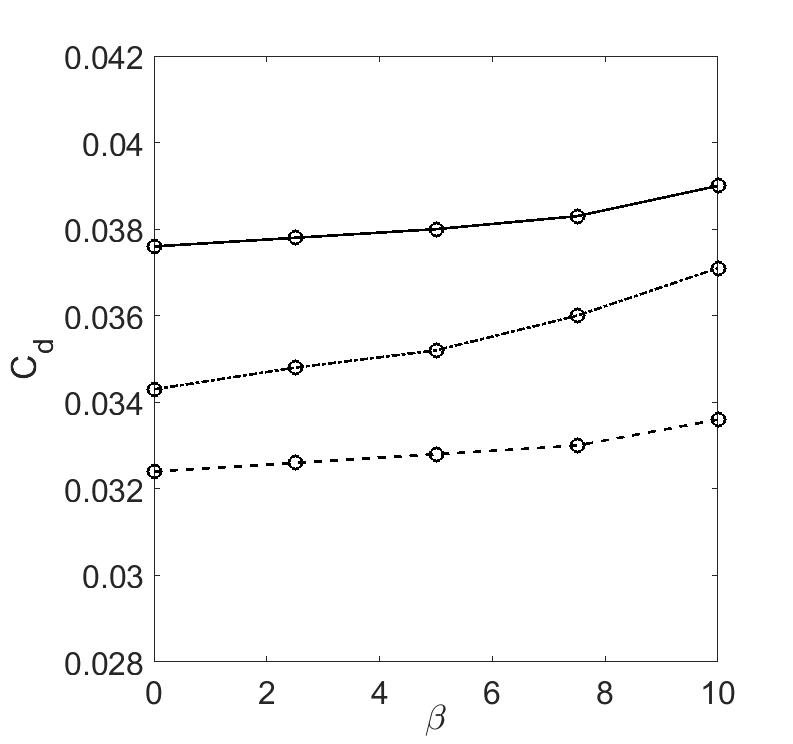}}
\subfloat[ ]{\includegraphics[height=8cm]{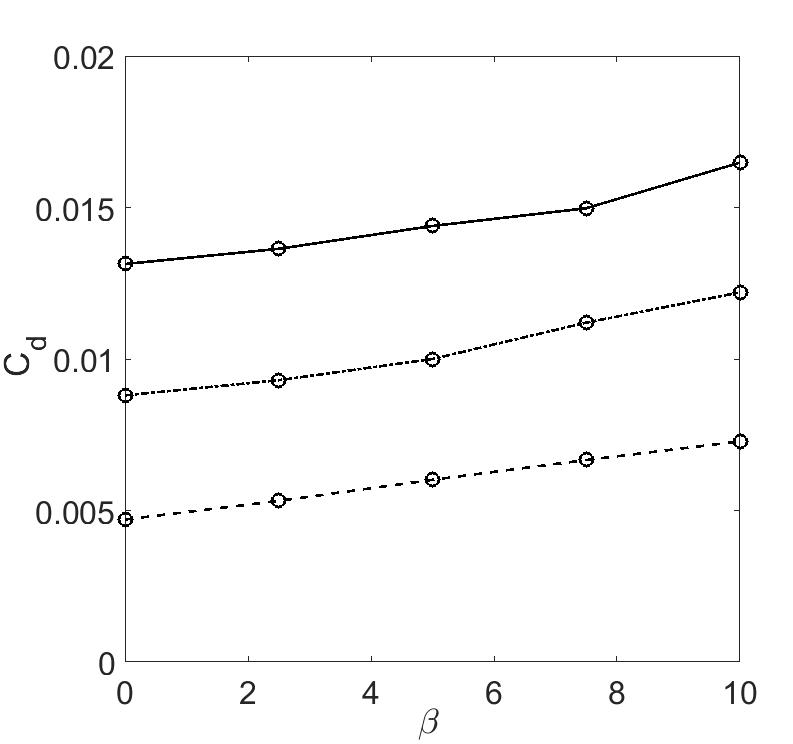}}
 \caption{The variation of drag coefficient for different values of DA($\beta$) a) with Reynolds number variation, dashed line, dashed-dot line and solid line represent $Re_v=1.31\times 10^5$, $Re_v=1.11\times 10^5$ and $Re_v=0.89\times 10^5$ respectively, b) with $h^*$ variation dashed line, dashed-dot line and solid line represent $h^*=0$, $h^*=0.5$ and $h^*=1$ respectively.\label{fig:12}}
 \end{figure}

\begin{figure}
\centering
\subfloat[ ]{\includegraphics[height=8cm]{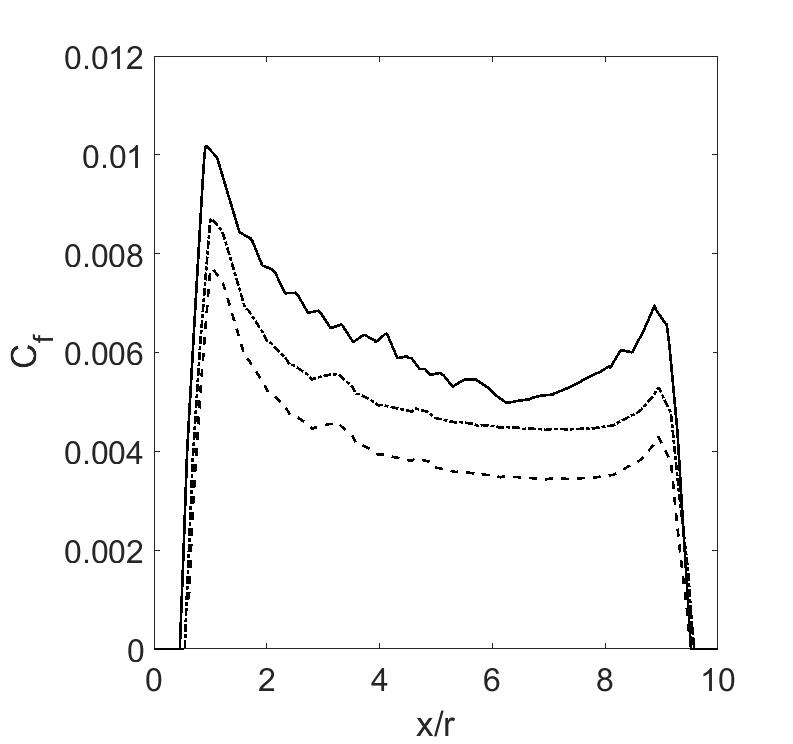}}
\subfloat[ ]{\includegraphics[height=8cm]{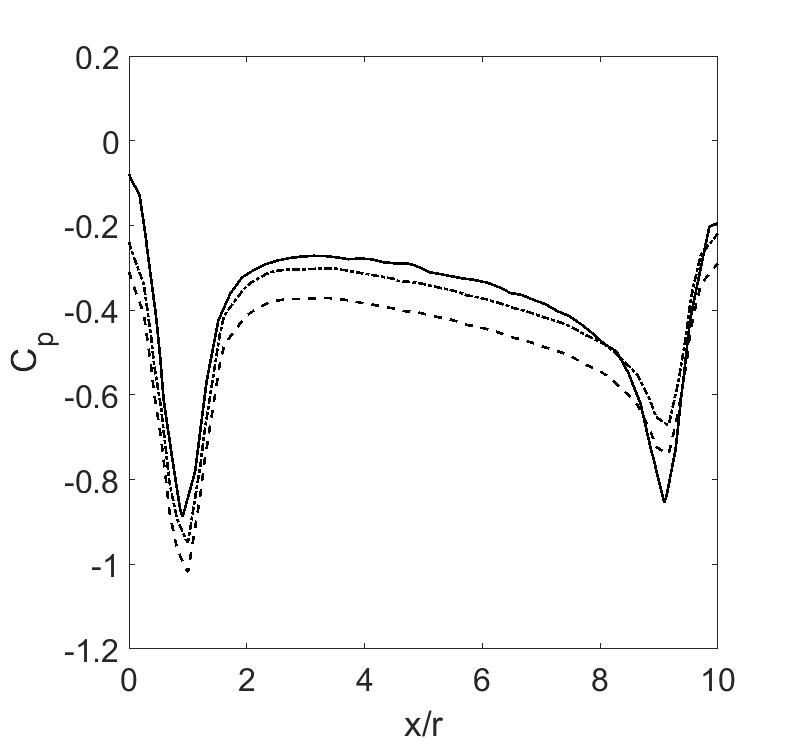}}
 \caption{The variation of skin friction and pressure coefficient for different values of DA, dashed line, dashed-dot line and solid line represent DA 0, 5 and 10 degrees respectively. \label{fig:13}}
 \end{figure}

%------------------------------------------------------------
In figure \ref{fig:12} the drag coefficient evolution for different values of drift angle are presented. The variation of drag coefficient with $\beta$ is presented for three different values of Reynolds numbers is presented in figure \ref{fig:12}a . It is noticed that with increase in drift angle, the drag coefficient increases. The same trend was preserved for all three values of volumetric Reynolds numbers. Figure \ref{fig:12}b represent the variation of drag coefficient for with $\beta$ for different wedge height. With increase in wedge height the drag coefficient is increased, the similar trend was also observed for drag coefficient variation with angle of attack. The skin friction and pressure coefficient evolution for different values of drift angle is presented in figure \ref{fig:13}. With increase in drift angle both skin friction and pressure coefficient increases along the AUV hull. The trend of pressure coefficient evolution for different values of drift angle is opposite that of the angle of attack. 

\newpage
\section{Concluding remarks}
Improvements in the design and testing of AUVs are becoming increasingly important due to their manifold applications in academic and industrial spheres. However, a majority of this testing and design is carried out under conditions that may not reflect the operating conditions for AUVs meant to operate in shallow depths. Such AUVs are deployed in coastal environments like the continental shelf and continental slope, along with estuaries. This difference in the design and testing conditions and the operating conditions may lead to imprecise estimations of the AUV's performance and sub-optimal designs. 

This study reports results of hydrodynamic coefficients evolution along an AUV hull at three different test bed slopes, using high fidelity RSM based simulations in conjunction with experiments. The experimental results of turbulent stresses and pressure coefficients were used to validate the Reynolds stress model predictions. From the RSM simulations it is observed that the hydrodynamic coefficients were very responsive to test bed slope variation. For example, small increases in test bed slope caused the drag coefficient to increase significantly. This has a cascading effect on the drag force as well. For instance, at a mild slope of $13^o$, the drag force on the AUV is over twice that in the flat test bed case. A critical comparison of flow evolution along the AUV hull for different angles of attack and drift angle was also performed. The drag coefficient increase with angles of attack and drift angle for all Reynolds number and wedge height. The experimental and numerical results presented in this article offer avenues for design improvement of AUVs operating in shallow environments, such as the coastal continental slope, the continental shelf and in estuaries. 

%%%%%%%%%%%%%%%%%%%%%%%%%%%%%%%%%%%%%%%%%%%%%%%%%%%%%%%%%%%%%%%%%%%%%%



%% References

\newpage
\section*{References}
\bibliographystyle{elsarticle-num}
\bibliography{asme2e.bib}






% \end{thebibliography}

\end{document}